# Fracture resistance of zigzag single walled carbon nanotubes

Qiang Lu[a] & Baidurya Bhattacharya[b]

[a]*Department of Mechanical Engineering, Northwestern University, Evanston, IL 60208, USA*

[b]*Department of Civil Engineering, Indian Institute of Technology, Kharagpur, WB 721302, India*

## Abstract

Carbon nanotubes (CNTs) are known to possess extraordinary strength, stiffness and ductility properties. Their fracture resistance is an important issue from the perspective of durability and reliability of CNT-based materials and devices. According to existing studies, brittle fracture is one of the important failure modes of Single-Walled Carbon Nanotube (SWNT) failure due to mechanical loading. However, based on the authors' knowledge, the fracture resistance of CNTs has not been quantified so far. In this paper, the fracture resistance of zigzag SWNTs with preexisting defects is calculated using fracture mechanics concepts based on atomistic simulations. The interatomic forces are modeled with a modified Morse potential; the Anderson Thermostat is used for temperature control. The problem of unstable crack growth at finite temperature, presumably caused by lattice trapping effect, is circumvented by computing the strain energy release rate through a series of displacement-controlled tensile loading of SWNTs (applied through moving the outermost layer of atoms at one end at constant strain rate of $9.4 \times 10^{-4}$/ps) with pre-existing crack-like defects of various lengths. The strain energy release rate, $G$, is computed for (17,0), (28,0) and (35,0) SWNTs (each with aspect ratio 4) with pre-existing cracks up to 29.5Å long. The fracture resistance, $G_c$, is determined as a function of crack length for each tube at three different temperatures (1K, 300K and 500K). A significant dependence of $G_c$ on crack length is observed reminiscent of the rising R curve behavior of metals at the macroscale: for the zigzag nanotubes $G_c$ increases with crack length at small length, and tends to reach a constant value if the tube diameter is large enough. We suspect that the lattice trapping effect plays the role of crack tip plasticity at the atomic scale. For example, at 300 Kelvin, $G_c$ for the (35,0) tube with aspect ratio 4 converges to 6 Joule/m$^2$ as the crack length exceeds 20 Angstrom. This value is comparable with the fracture toughness of graphite and Silicon. The fracture resistance of the tubes is found to decrease significantly as the temperature increases. To study the length effects, the computations are repeated for zigzag nanotubes with the same three chiralities but with aspect ratio 8 at 1K. The fracture resistance of the longer nanotubes are found to be comparable to those of the shorter nanotubes.

## Keywords

## 1. Introduction

Carbon nanotubes (CNTs) are one or more layers of helical carbon microtubules, in which each layer can be described as rolling a single sp$^2$ graphene sheet into a cylinder along a vector called the chiral vector (*m,n*). Although earlier findings related with tubular shape of carbon atoms had been reported, it was Iijima's report [1] that gave rise to the current wave of enthusiasm on CNTs.

CNT can be classified as single-walled nanotube (SWNT) and Multi-Walled nanotube (MWNT). MWNTs usually have diameters of less then 100nm, and length in micrometers [1-4]. CNTs have been found to possess excellent electronic, mechanical, optical and thermal properties, which make the carbon nanotube a potentially very useful material and CNTs are now used as fibers in composites, scanning probe tips, field emission sources, electronic actuators, sensors, Li ion and hydrogen storage and other electronic devices. Nowadays, various techniques, such as arc-discharge, laser-ablation, and catalytic growth are applied to produce CNTs.

CNTs have extraordinary mechanical properties that have been the subject of both computational and experimental investigations since the early 1990s. For example, the CNT Young's modulus was measured through thermal vibration [5,6] and resonant vibration [7,8] tests of cantilevered CNTs, by direct stretching tests [9,10] of CNT samples, and by bending/buckling tests [11,12] using atomic force microscopy. These experiments suggested that CNT Young's modulus is usually around 1 TPa, and single-walled nanotubes (SWNT) is slightly stiffer than multi-walled nanotubes. Experiments also show that Young's modulus drops quickly as the diameter of tubes increases [5,7,8,13]. Tensile strength of CNTs obtained from direct stretching tests [9,10,14] is found to vary between 10 and 150 GPa. The Poisson's ratio of CNTs, is found to vary from 0.14 to 0.34 according to various simulation studies [15]. Radial deformation of CNTs has been investigated and it was found that CNTs are very resilient and can be reversibly deformed up to a strain of 50% [16,17]. CNTs were also found to be robust under hydrostatic pressure. Experiments reported that SWNTs were reversibly compressed to a high pressure (2.9 GPa reported by [18], 4 GPa by [19], or even 10 GPa by [20]). Twisted and collapsed CNTs were observed with TEM [21,22], and the highly deformed structures seemed to be quite stable. Bending tests [7,11,23,24] showed that CNTs are extraordinarily flexible under large strains. CNTs can be bent to large angles without apparent distortion in local structures. Beyond certain angles, local buckling happens at the compression side of the tube, but is generally elastic and reversible. The MWNT walls can even flatten into deflated ribbons under torsion but still can restore their original shape. The capability of non-linear but reversible local deformation provides CNTs excellent resilience in bending and compression [14,25]. The bending strength of MWNTs has been obtained as 14.2 +/- 8.0 GPa using the strain at the initial buckling point by Wong et al [11]. Shear and interlayer sliding of MWNTs were also studied with experiments [26-28].

The fracture resistance of CNTs could be a key factor when they are used in load-bearing or load-sensing materials/devices and must be properly quantified before such CNT-based products are adopted for widespread use. However, to the knowledge of the authors, the fracture resistance of CNTs have not be quantified with any precision thus far.

Presumably due to their small size, few experimentally observed details have been reported in the literature on the fracture properties and processes of CNTs. In a series of high-resolution TEM and tight binding simulation studies [29,30] on fracture of CNTs, the CNT sample was formed by thinning a carbon film with the aid of an intense electron beam. This thinning mechanism turns out to be an efficient method to release the axial tension in the fiber. It was reported that both brittle and ductile fracture could be possible for SWNTs. However, under what conditions the SWNT would favor brittle/ductile fracture modes was not explained by the experiment.

Based on atomistic simulations, another study [31] found that SWNTs can also undergo plastic deformation due to mechanical loading, depending on the external conditions and tube symmetry. The plastic deformation is believed to result from successive formation of Stone-Wales (SW) rotations and gradually changing of CNT configurations. However, the above explanations of plastic deformation were challenged by two other studies. Dumitrica et al. [32] argued that the energy barrier for SW defects formation at room temperature is high enough to inhibit stress-induced SW defects formation, and showed the direct bond-breaking is a more likely failure mechanism for defect-free nanotubes. In the same study [32], by modeling the energy states of SWNTs with electronic structure programs, they showed that brittle fracture is the dominant failure mode at low temperature. A quantum mechanical study [33] pointed out that the results in [31] and related works were not reliable since the potential model they used (i.e., the Bond Order model) was not capable of correctly describing breaking of the bond connecting the two pentagons in the SW pair. That same study [33] further found that pre-existing SW defects caused successive bond breakings instead of bond rotations as reported by [31,34-36]. Based on above studies, some general conclusions can be drawn: zigzag tubes favor brittle mode, while armchair tubes, under high temperature and high strain, can sometimes undergo significant plastic deformation before fracture.

Further, it is well-known that defects are commonly present in nanotubes (either resulting from the manufacturing process or intentionally introduced to improved functionalities): vacancies, metastable atoms, pentagons, heptagons, Stone-Wales defects, heterogeneous atoms, discontinuities of walls, distortion in the packing configuration of CNT bundles, etc. are widely observed in CNTs [37-40]. According to an STM observation of the SWNT structure, about 10% of the samples were found to exhibit stable defect features under extended scanning [41]. Defects can also be introduced by mechanical loading [31] and electron irradiation [37]. These defects have been found to have significant influence on CNT mechanical properties [42,43]. It is reasonable to assume that such defects could initiate brittle fracture in CNTs under appropriate conditions. In this paper, the authors try to investigate the brittleness of zigzag SWNTs with pre-existing defects using metrics commonly used in fracture mechanics based on displacement- controlled constant-temperature loading of the tubes through atomistic simulation.

## 2. Atomistic simulation of SWNTs under mechanical loading

Atomistic simulations, if performed correctly, provide an efficient and economical alternative to actual laboratory experiments in studying small scale mechanical processes. In this study, the carbon-carbon interaction is modeled with a modified Morse potential [44] that has been applied to study SWNT mechanics. The potential energy has the form:

$$E_i = E_{stretch} + E_{angle} = \sum_j E_{stretch}(i,j) + \sum_{jk} E_{angle}(i,j,k) \tag{1}$$

$$E_{stretch}(i,j) = D_e\{[1 - e^{-\beta(r-r_0)}]^2 - 1\} \tag{2}$$

$$E_{angle}(i,j,k) = \frac{1}{2}k_\theta(\theta_{ijk} - \theta_0)^2[1 + k_{sextic}(\theta_{ijk} - \theta_0)^4] \tag{3}$$

This is the usual Morse potential except that the bond angle-bending energy has been added and the constants are slightly modified so that it corresponds with the Brenner potential [45] for strains below 10% [44]. $E_{stretch}$ in Eq (1) is the potential energy due to bond strength, $r$ is the length of the bond. $E_{angle}$ in Eq (1) is the potential energy due to the bond angle bending, $\theta$ is the current angle of the adjacent bonds. The potential model parameters are $r_0 = 1.39 \times 10^{-10}$ m, $D_e = 6.03105 \times 10^{-19}$ Nm, $\beta = 2.625 \times 10^{10}$ m$^{-1}$, $\theta_0 = 2.094$ rad, $k_\theta = 0.9 \times 10^{-18}$ Nm/rad$^2$, $k_{sextic} = 0.754$ rad$^{-4}$. For computational convenience, time, distance and quantities representing velocity, energy etc. are reduced to non-dimensional numbers during the simulation. Table 1 shows the reduction of units.

The bond order model is frequently used in simulations of carbon nanotubes. However, as pointed out by a few studies [33,44,46], the cutoff function in Brenner's bond order model is said to give rise to spurious forces and inaccurately large breaking strain. As discussed in Section 1, the fracture modeling using bond order model gave different fracture mechanism from the studies of *ab initio* calculation. The Modified Morse model seems to perform better in the fracture properties [44]. However, studies have shown that correctly reproducing the phonon dispersion properties of a crystal is important for thermodynamic stability of the lattice structure and the simulation of the thermal and electronic properties of the crystal. According to existing studies, up to the fourth closest neighbor interactions are need for correctly reproducing the phonon dispersion properties of graphene [47]. However, because of the curvature of carbon nanotube, the calculation of phonon modes required different interaction terms for consideration of inclusion of force constants [39]. Since the Modified Morse potential considers only the $1^{st}$ and $2^{nd}$ closest neighbors, there is a possibility that it may not be sufficient for correctly reproducing the phonon dispersion properties of carbon nanotubes. However, the calculation of phonon modes with the Modified Morse potential could not be located in the literature by the authors. Nevertheless, the Modified Morse potential has already been shown in literature [44] and our benchmark simulations [48] to be able to keep thermodynamic stability of the nanotube, and to reproduce carbon nanotube mechanical properties (such as elastic modulus and strength) fairly well. Therefore, since this paper concerns the fracture resistance of carbon nanotubes, the authors consider phonon dispersion properties as less of a focus, and adopt the Modified Morse potential for atomistic simulations in this paper.

Figure 1 (a) shows a (17,0) tube of aspect ratio 4 (length, $l$ = 23.4 Å) that is loaded to failure at room temperature (300 Kelvin). The time histories of energies, temperature, force and displacement until the tube breaks are shown in part (b) of the Figure. The tube is stretched by forcing the atoms at one end to move at constant speed. The loading rate is $v$ = 1.67×10$^{-4}$ r.u. (reduced unit), which produces a displacement of 5.0 nm in 1 ns. The initial atomic velocities are randomly chosen according to a uniform distribution (between the limits −0.5 and 0.5) and then rescaled to match the initial temperature of 300 Kelvin. At the beginning of the simulation, the tube is relaxed for about 200 reduced units.

The tube is found to exhibit brittle behavior at fracture: once the tube deformation reaches a critical level, atomic bonds break successively and lead to a complete fracture in a very short time. The detailed fracture process starting at around time 2256 r.u. and ending around time 2263 r.u. is shown in Figure 2 (a) while the corresponding force time history is magnified in Figure 2 (b). A total of 8 snapshots marked A through I have been identified and reproduced in Figure 2 (a), each of which corresponds to the breaking of one or more atomic bonds. The same 8 points have also been indicated in the time history of Figure 2 (b).

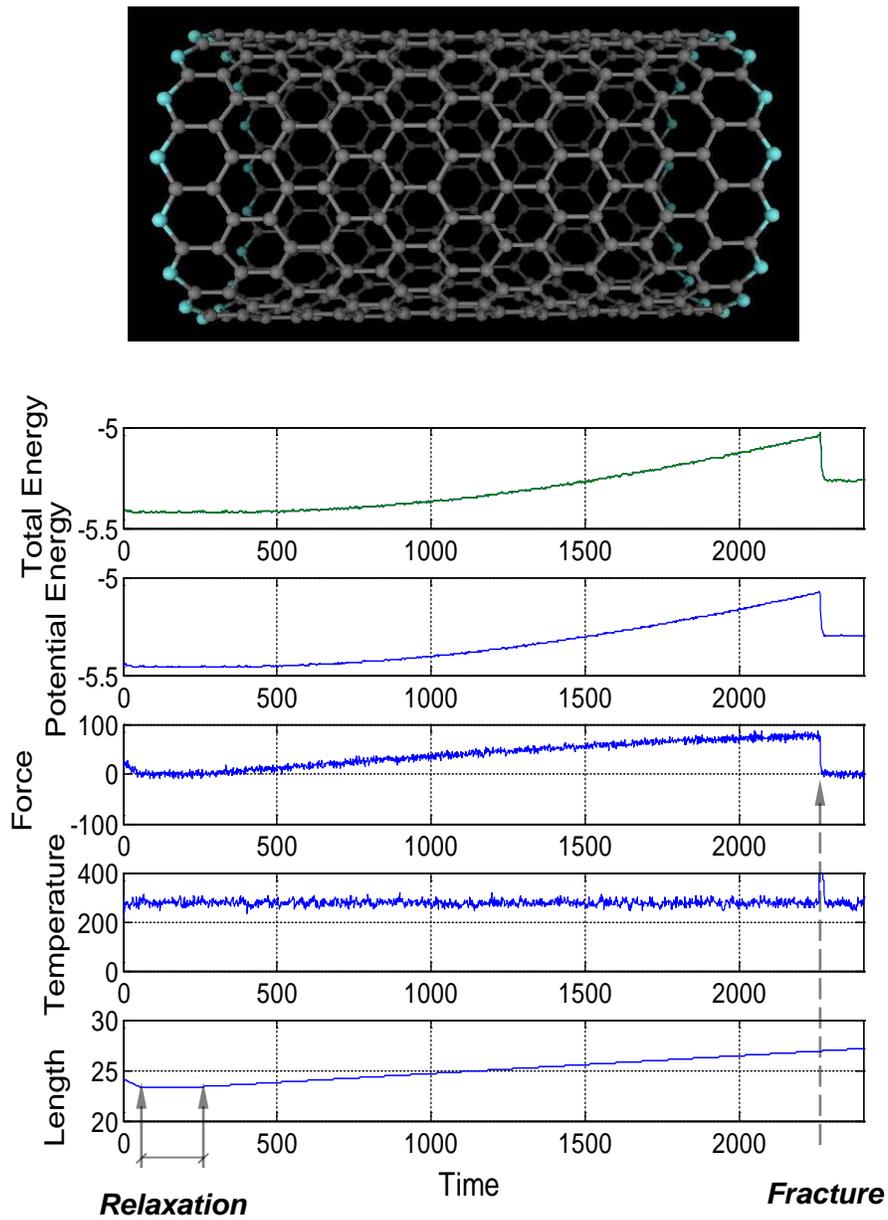

**Figure 1 (a) A (17,0) zigzag SWNT with length 23.4 Å (highlighted atoms are subjected to displacement controlled loading at 300K), and (b) Atomistic simulation of the time histories to fracture**

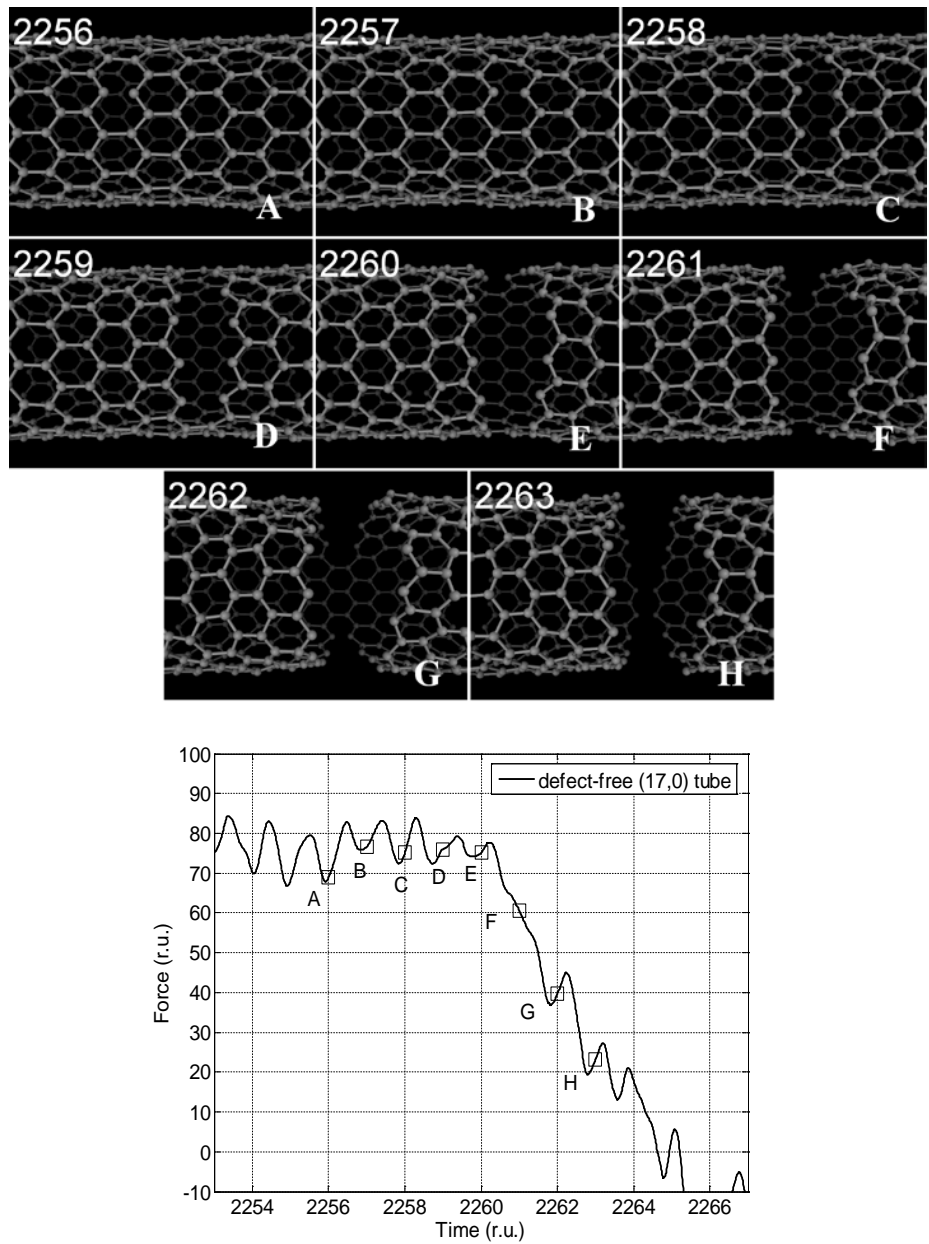

**Figure 2 Detailed fracture process for a (17,0) SWNT with length 23.4Å; (a) (A-H) Snapshots (b) corresponding force time history**

The time histories in Figure 1 (b) can be used to verify some of the well-known mechanical properties of SWNTs. The axial force $F_z$ is calculated by summing the forces acting on the atoms in the axial direction at one end of the tube, and stress is calculated as $\sigma_{zz} = F_z / A$ in the following computations. The cross-sectional area, $A$, of the tube is taken to be 139.1 Å$^2$, based on the commonly adopted value of tube wall thickness of 0.34 nm [2] as stated above. The deformation of SWNT is not purely linear, the stiffness decreases gradually up to the break point, as shown in Figure 3 (a). Part (b) of the figure adopts this force-displacement relation in computing strain energy release rates for fracture resistance measurement as discussed in detail in Section 3.

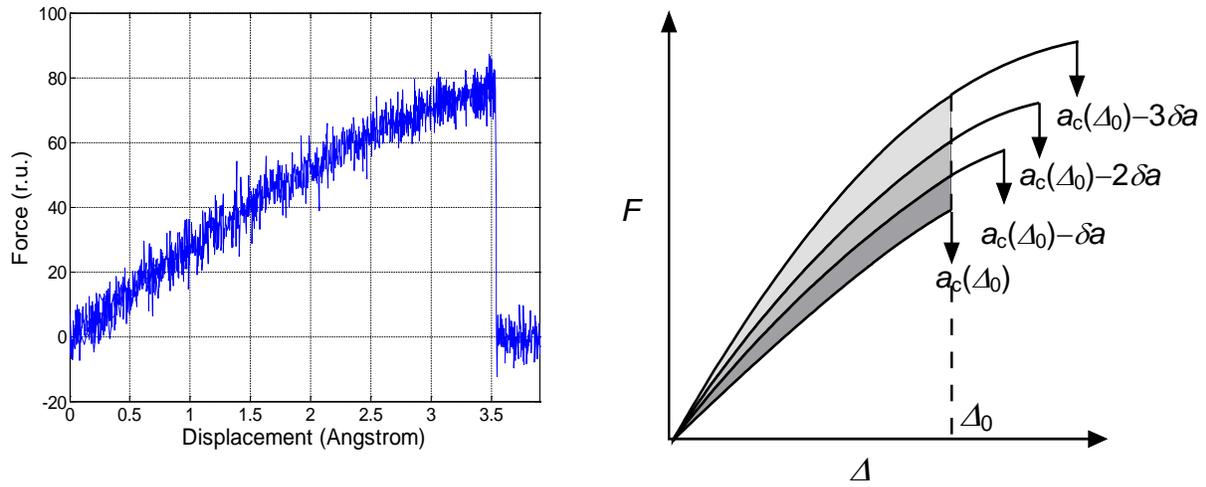

**Figure 3. (a) Force-displacement curve of the (17,0) SWNT with length 23.4Å (b) Method used in this study for determining strain energy release rate using displacement-controlled fracture test**

**Table 1 Reduced units for atomistic simulation**

| Quantity | Reduced units | Real units |
|---|---|---|
| Length | 1 | $1 \times 10^{-10}$ m |
| Energy | 1 | eV = $1.602 \times 10^{-19}$ J |
| Mass | 1 | $1.992 \times 10^{-26}$ kg |
| Temperature | 1 | $1.1609 \times 10^{4}$ K |
| Time | 1 | $3.526 \times 10^{-14}$ s |
| Force | 1 | $1.602 \times 10^{-9}$ N |
| Pressure | 1 | 160.2 GPa |
| Speed | 1 | $2.836 \times 10^{3}$ m/s |

Common mechanical properties can be calculated from the simulated force and displacement time histories above. For example, (i) The Young's modulus can be calculated as the initial slope of the force-displacement curve (multiplied with $l/A$) after a second order polynomial fit is applied to the force-displacement curve. The resulting value for the (17,0) SWNT at 300K is $E = 0.98$ TPa. (ii) The ultimate strength can be calculated at the maximum force point, $\sigma_u = F_{max}/A$, where $F_{max}$ is the maximum axial force. The computed value is $\sigma_u = 98.1$ GPa. (iii) The ultimate strain, which is the strain corresponding to the ultimate strength, can be calculated as $\varepsilon_u = \Delta L_u/l$. The engineering definition of strain is adopted for this example (i.e., with respect to the original undeformed dimensions). The ultimate strain is found to be 11.58%. (iv) The Poisson' ratio can be calculated as the ratio between the strain in the lateral direction (averaged in two perpendicular directions at the midsection) and the strain in the axial direction. In this example involving the (17,0) SWNT at 300K, the Poisson's ratio is found to be 0.238.

### 3.    Fracture resistance analysis of zigzag SWNTs

Elegant fracture theories have been developed during the past decades as reviewed in Lu and Bhattacharya [48] for fracture of materials at macro-scales.  It is important to note here that most of these approaches are continuum based.  As reviewed in [48], continuum approaches may not always be appropriate when details at the atomic scale become important. Concepts such as stress, density, crack surface area, stress intensity factor etc. are all based on continuum assumptions which do not necessarily hold at the atomic scale.  For example, the stress field in the absence of homogenization amounts to a discrete field with singularities at the atomic scale.  Since energy has the same meaning at all scales in thermomechanics [49], we adopt an energy-based approach to fracture in this paper, rather than one that is stress based.

Based on the stress solution proposed by Inglis [50], Griffith [51] treated fracture as an equilibrium process in which the loss of strain energy can be equated to the surface energy generated due to the growth of cracks.  Irwin [52] introduced the parameter strain energy release rate, $G$, as a measure of the energy available for crack extension (also referred to as the crack driving force).  Consider a solid body in equilibrium containing a crack with length, $a$.  For the crack to extend, enough energy must be provided to overcome the surface energy of the material.  The critical condition of energy balance for an incremental increase in crack area is

$$\frac{d\Pi}{dA} + \frac{dW_s}{dA} + \frac{dE_K}{dA} = 0 \qquad (4)$$

where $\Pi$ is the potential energy, $E_K$ is the kinetic energy, $W_s$ is the work required to create new surfaces, and $A$ is the crack surface area.  The strain energy release rate, $G = -d\Pi/dA$ equals $G_C$ at the onset of crack growth; $G_C$ is a measure of fracture resistance.   For brittle materials like glass, $G_C$ is a (temperature-dependent) material property independent of crack length.  On the other hand, ductile materials which allow plastic deformation ahead of the crack tip typically exhibit $G_C$ that increases with crack-length up to a point (and depends also on specimen geometry to some extent) – this property allows for stable crack growth in metals and is known as "rising R curve" (i.e., resistance curve) behavior [53].

Based on the force-displacement curve obtained through atomistic simulation, the work done by the external force (let us denote this work by $W$) can be calculated by integrating the area under this curve.  Since temperature control is applied, the kinetic energy part $dE_K/dA$ is negligible compared to the potential part.  Because the only net energy input is the work done by the external force, the net change of potential energy $\Pi$-$\Pi_0$ ($\Pi_0$ is the potential energy of the system at equilibrium) approximately equals $W$ up to the onset of fracture, and hence we assume  $d\Pi/dA \approx dW/dA$  in the following.

The method used in this study to calculate the fracture resistance is based on the displacement-controlled test method commonly used for macro scale fracture specimens.  However, important modifications are needed to account for the dynamics at the atomic level before this method can be applied to CNTs, as described in the following.  At finite temperatures, fracture happens in a catastrophic manner in all our simulations.  That is presumably because of the lattice trapping effect [32], which basically makes the energy barrier higher than what is needed for creating new surface.  The extra part of energy is thus converted into kinetic energy, which causes a sudden increase in the kinetic energy (as shown in Figure 1 (b)), and helps the local atoms on the edge of crack overcome the energy barrier for the crack to extend further.  Thus, it is difficult to obtain a stable crack growth at fixed displacement with atomistic simulation.

Therefore, instead of running one simulation for crack growth from $a$ to $a+\delta a$, then to $a+2\delta a$ and so on, up to $a_c(\Delta_0)$ (where $a_c(\Delta_0)$ is the crack length at the onset of catastrophic failure at fixed displacement $\Delta_0$), we run several separate simulations of tensile loading of SWNTs to fracture, with pre-existing crack lengths $a_c(\Delta_0)-n\delta a$, … , $a_c(\Delta_0)-3\delta a$, $a_c(\Delta_0)-2\delta a$, $a_c(\Delta_0)-\delta a$ and $a_c(\Delta_0)$, respectively. The smallest crack length, $\delta a$, in the circumferential direction is the distance between two nearest longitudinal bonds, i.e. $\delta a = 2.46$ Å. Here $n = (a_c - \delta a)/\delta a$; $n$ is an integer because the nanotube wall is composed of discrete network of carbon atoms. The strain energies $W_n$, …, $W_3$, $W_2$, $W_1$ and $W_0$, at the fixed displacement $\Delta_0$, can be determined from the force-displacement curves, as shown in Figure 3 (b). Thus a $W$~$a$ curve can be drawn for this fixed displacement $\Delta_0$ as shown in Figure 4. This curve in turn can be used to calculate the slope $dW/dA$ (and hence its negative the crack driving force, $G$) at various crack lengths. In particular, at the critical crack length $a_c(\Delta_0)$ the driving force equals the resistance, $G_c$. Since $a_c(\Delta_0)$ depends on the fixed displacement, $\Delta_0$, the resistance, $G_c$, can be computed for different $\Delta_0$ values and hence for different crack lengths. This way, $G_c$ vs. $a$ curves can be obtained at various temperatures for different tube configurations.

## 4. Discussion of Results

We choose 3 zigzag SWNT configurations: (17,0), (28,0) and (35,0), each with an aspect ratio of 4 (giving lengths of 23.4, 39.1 and 50.0 Angstroms respectively), at temperatures ranging from 1 to

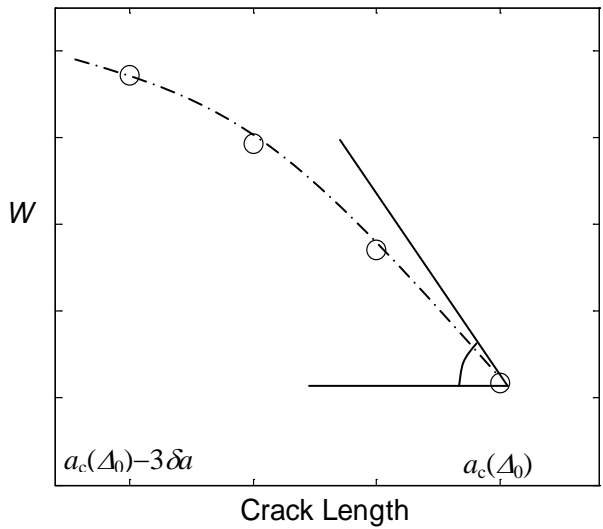

500K. Tensile loading is applied to nanotubes with 0, 1, 2, ..., $n$ bonds cut at a time. The values of $n$ are 4, 7 and 11, respectively, for the (17,0), (28,0) and (35,0) nanotubes. The initial crack length for all nanotubes is set to be $\delta a = 2.46$ Å. The constant loading speed is 5.0 nm/ns, 8.5 nm/ns and

9.8 nm/ns, respectively, for the (17,0), (28,0) and (35,0) SWNTs, so that the strain rate is constant ($9.4 \times 10^{-4}$/pico second). Temperature is controlled with Anderson thermostat in which the probability of collision for each atom in each time step is set at 0.005.

**Figure 4 External work-Crack length (W-a) curve for SWNTs**

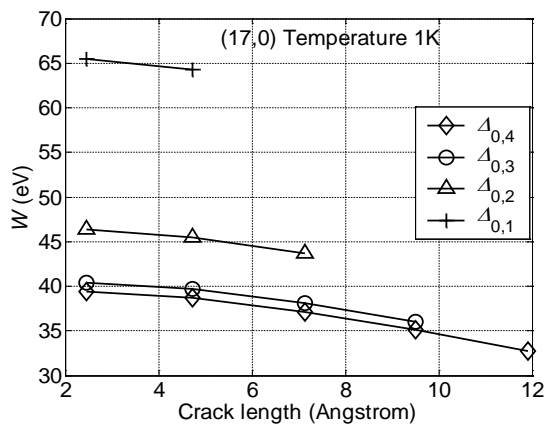
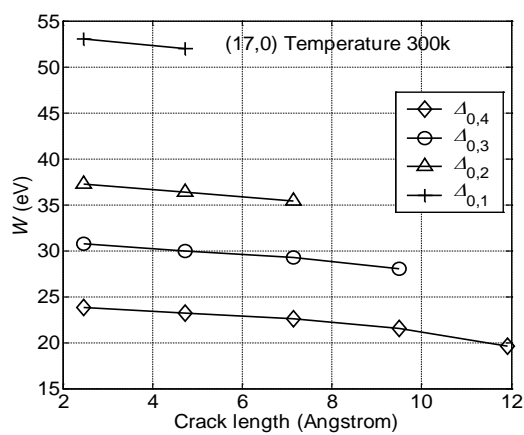

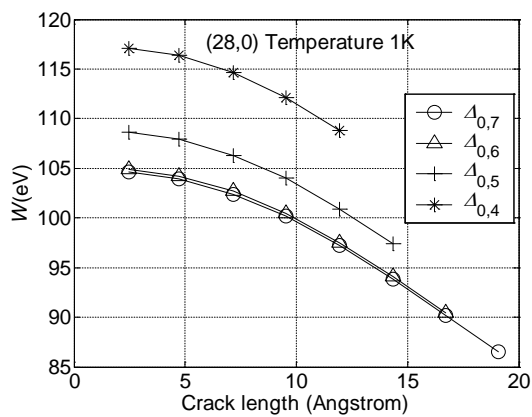
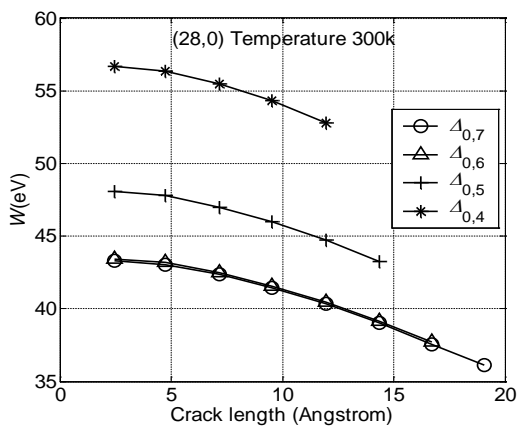

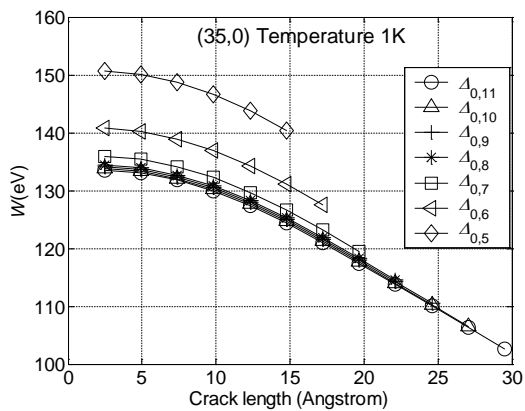
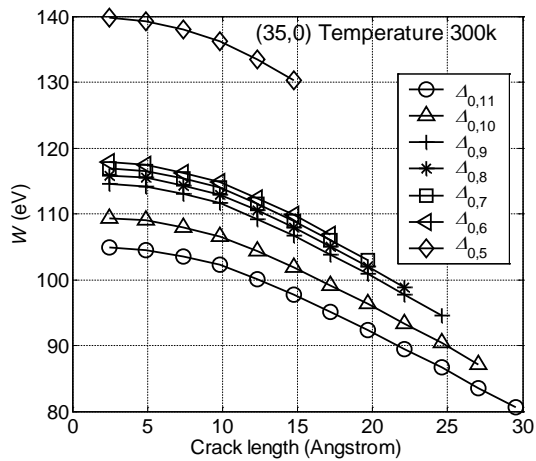

**Figure 5 External work-crack length (W-a) curve at 1 K for (a) (17,0), (b) (28,0) and (c) (35,0) SWNTs with aspect ratio 4**

**Figure 6 External work-crack length (W-a) curve at 300 K for (a) (17,0), (b) (28,0) and (c) (35,0) SWNTs with aspect ratio 4**

Figure 5 (a) shows the *W-a* data for the (17,0) zigzag SWNTs at temperature 1K at different fixed displacements, $\Delta_{0,i}$ ($\Delta_{0,i}$ is the displacement at which the critical crack length $a_c(\Delta_{0,i}) = 2.46(i+1)$ Å). The terminal slope of each curve gives $G_c$ at the respective critical crack length. It can be seen that the terminal slope varies with *i*, suggesting crack-length dependent fracture resistance. This aspect is probed in detail subsequently.

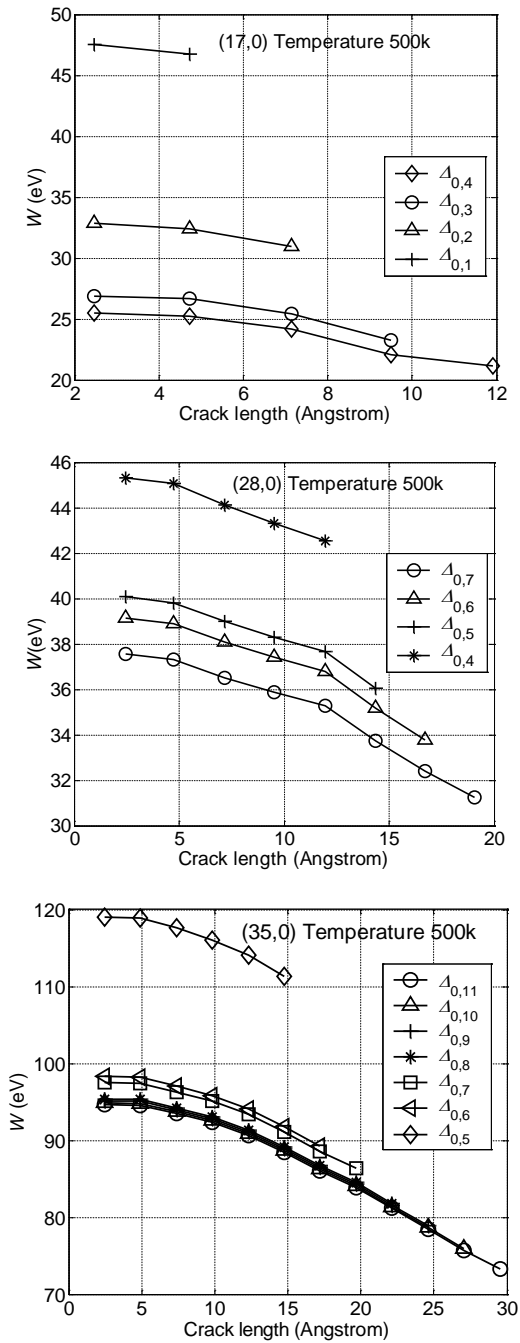

**Figure 7 External work-crack length (W-a) curve at 500 K for (a) (17,0), (b) (28,0) and (c) (35,0) SWNTs with aspect ratio 4**

For nanotubes with larger diameters, it is possible to investigate the fracture properties with longer cracks. Using the same method as above, (28,0) and (35,0) SWNTs with 0~7 bonds cut, and 0~11 bonds cut respectively (maximum crack length 19.68 and 29.52 Å, respectively) are studied. Their *W-a* relations at temperature 1K are shown in Figure 5 (b) and (c). We see that *W-a* curves have a tendency to converge at higher displacements if the tube is wide enough.

The effects of temperature on *W-a* plots are also investigated for each tube. We choose two higher temperatures: 300 and 500 Kelvin. Corresponding *W-a* plots are shown in Figure 6 and Figure 7.

We now turn our attention to estimating fracture resistance of the three tubes at the three different temperatures. As stated above, $G_c$ is given by the negative of the right terminal slope of each *W-a* curve; the slope is computed using 3-point backward finite difference method. Based on the *W-a* data from the previous figures, $G_c$ vs. crack length for the three nanotube are shown in Figure 8 for 1, 300, 500 Kelvin respectively. A significant dependence of $G_c$ on crack length is observed. $G_c$ increases with crack length initially, and tends to reach a constant value, about 7.0 Joule/m$^2$ as crack length is large, as graphed in the Figure. Another interesting thing is, this constant value seems not to be related with a specific tube diameter. This behavior is strongly evocative of the rising R curve behavior for metals; we suspect that lattice trapping effect plays the role of crack-tip plasticity at the atomic level.

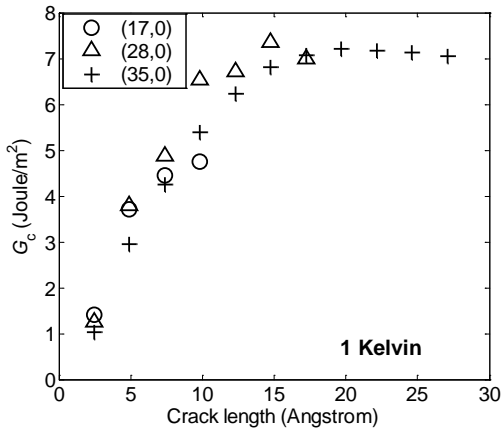

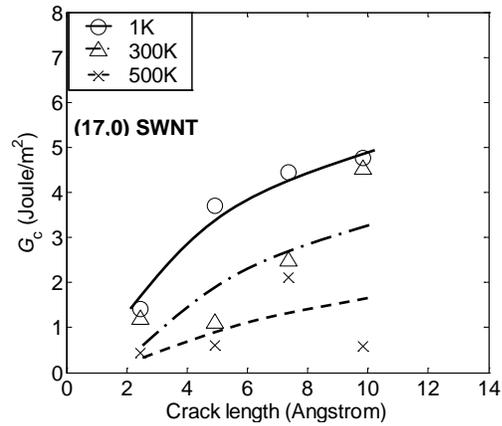

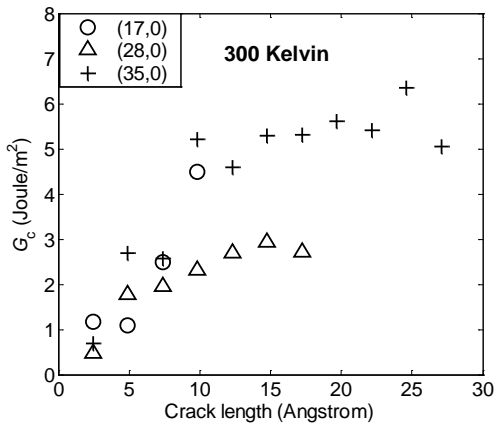

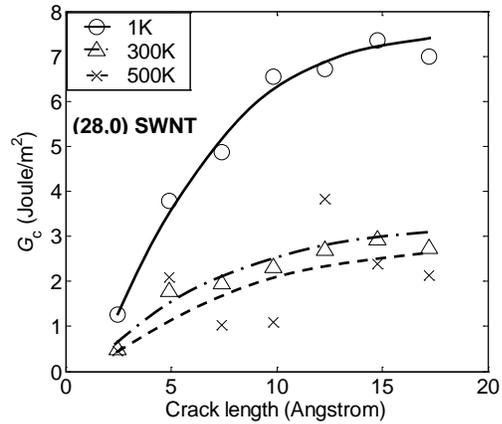

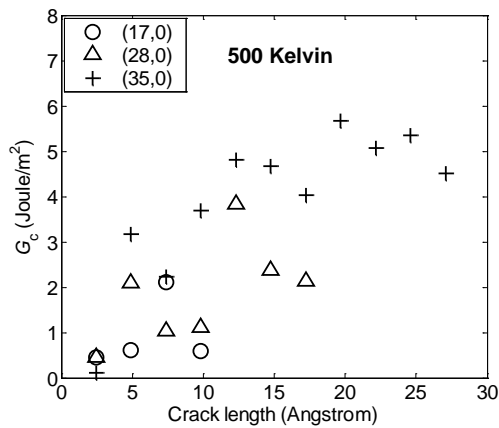

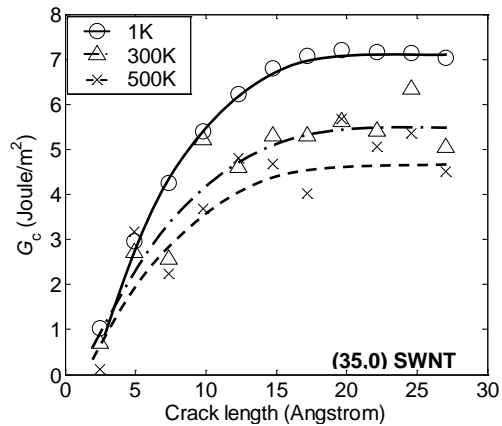

**Figure 8** $G_c$ (Joule/m$^2$) vs. length for (17,0), (28, 0) and (35,0) SWNT with aspect ratio 4 at (a) 1, (b) 300 and (c) 500 Kelvin

**Figure 9** Critical strain energy release rate at 1, 300 and 500 K for (a) (17,0) SWNT (b) (28,0) SWNT (c) (35,0) SWNT with aspect ratio 4

The dependence of $G_c$ on temperature is also significant. At higher temperatures, *W-a* plots have increased fluctuations, which makes it more difficult to compare with each other. From these *W-a* data, it is found that $G_c$ drops substantially as temperature increases, and the trend is graphed in Figure 9 for (17,0), (28,0) and (35,0) SWNTs. For example, at large crack lengths, the asymptotic value of $G_c$ drops from 7.0 to 5.5 and then to 4.9 J/m$^2$ at 1, 300 and 500 K, respectively. However, the flattening trend of $G_c$ vs. $a$ curve is retained.

The values of $G_c$ at the different crack length for 1, 300 and 500 Kelvin for nanotubes of the three different diameters are summarized in Table 2.

**Table 2** $G_c$ **(Joule/m$^2$) at different critical crack length $\Delta_0$ for (17,0), (28, 0) and (35,0) SWNT with aspect ratio 4 at 1, 300 and 500 Kelvin**

| Crack length, $a$ (Å) | Fracture resistance, $G_c$  (J/m$^2$) | | | | | | | | |
|---|---|---|---|---|---|---|---|---|---|
| | (17,0) SWNT | | | (28,0) SWNT | | | (35,0) SWNT | | |
| | 1K | 300K | 500K | 1K | 300K | 500K | 1K | 300K | 500K |
| 27.06 | — | — | — | — | — | — | 7.044 | 5.049 | 4.503 |
| 24.60 | — | — | — | — | — | — | 7.134 | 6.341 | 5.353 |
| 22.14 | — | — | — | — | — | — | 7.163 | 5.402 | 5.064 |
| 19.68 | — | — | — | — | — | — | 7.212 | 5.605 | 5.678 |
| 17.22 | — | — | — | 6.987 | 2.720 | 2.126 | 7.072 | 5.304 | 4.023 |
| 14.76 | — | — | — | 7.358 | 2.927 | 2.378 | 6.810 | 5.297 | 4.680 |
| 12.30 | — | — | — | 6.706 | 2.683 | 3.824 | 6.237 | 4.596 | 4.810 |
| 9.84 | 4.759 | 4.499 | 0.5894 | 6.536 | 2.307 | 1.103 | 5.400 | 5.206 | 3.683 |
| 7.38 | 4.440 | 2.482 | 2.1148 | 4.861 | 1.950 | 1.038 | 4.245 | 2.564 | 2.233 |
| 4.92 | 3.703 | 1.094 | 0.6050 | 3.797 | 1.766 | 2.097 | 2.953 | 2.699 | 3.167 |
| 2.46 | 1.409 | 1.173 | 0.4425 | 1.256 | 0.476 | 0.453 | 1.0347 | 0.6866 | 0.111 |
| Note: — indicates $a$  is too large for given SWNT configuration, hence $G_c$ is not calculated | | | | | | | | | |

In order to investigate the effects of the length of the nanotubes, simulations are conducted for another set of carbon nanotubes with the aspect ratio twice that of the original set. Common tensile properties (the Young's modulus, the ultimate strength, the ultimate strain and Poisson's ratio) as well as fracture resistance at different crack lengths are computed. Because the fracture resistance data at 300 and 500 Kelvin has much fluctuations and thus hard to compare, only fracture resistances at 1 Kelvin are compared. The comparison between the two sets are shown below in Table 3 and

| | Mechanical properties | | | | | |
|---|---|---|---|---|---|---|
| | (17,0) SWNT | | (28,0) SWNT | | (35,0) SWNT | |
| Aspect ratio | 8 | 4 | 8 | 4 | 8 | 4 |
| Length (Å) | 48.2 | 23.4 | 81.2 | 39.1 | 101.8 | 50.0 |
| E (TPa) | 1.13 | 1.12 | 1.16 | 1.15 | 1.19 | 1.16 |
| $\sigma u$ (GPa) | 94.98 | 95.05 | 94.57 | 94.56 | 94.37 | 94.51 |
| $\varepsilon u$ | 14.96% | 15.07% | 14.60% | 14.66% | 13.75% | 14.61% |
| $\nu$ | 0.303 | 0.305 | 0.301 | 0.302 | 0.296 | 0.296 |

Table 4 respectively. The tensile properties appear to be quite insensitive to length changes. The fracture resistance shows no definite trend, although in a majority of the cases the longer nanotubes are found to have higher fracture resistance than the shorter tubes, but the difference seems to decrease as the tube diameter increases.

**Table 3 Comparison between mechanical properties of nanotubes with different aspect ratios, at 1 K**

| | Mechanical properties | | | | | |
|---|---|---|---|---|---|---|
| | (17,0) SWNT | | (28,0) SWNT | | (35,0) SWNT | |
| Aspect ratio | 8 | 4 | 8 | 4 | 8 | 4 |
| Length (Å) | 48.2 | 23.4 | 81.2 | 39.1 | 101.8 | 50.0 |
| E (TPa) | 1.13 | 1.12 | 1.16 | 1.15 | 1.19 | 1.16 |
| $\sigma_u$ (GPa) | 94.98 | 95.05 | 94.57 | 94.56 | 94.37 | 94.51 |
| $\varepsilon_u$ | 14.96% | 15.07% | 14.60% | 14.66% | 13.75% | 14.61% |
| $\nu$ | 0.303 | 0.305 | 0.301 | 0.302 | 0.296 | 0.296 |

**Table 4 Comparison between fracture resistances of nanotubes with different aspect ratios, at 1 K**

| Crack length, $a$ (Å) | Fracture resistance, $G_c$ (J/m$^2$) | | | | | |
|---|---|---|---|---|---|---|
| | (17,0) SWNT | | (28,0) SWNT | | (35,0) SWNT | |
| Aspect ratio | 8 | 4 | 8 | 4 | 8 | 4 |
| 27.06 | — | — | — | — | 8.8716 | 7.044 |
| 24.60 | — | — | — | — | 8.7717 | 7.134 |
| 22.14 | — | — | — | — | 7.3175 | 7.163 |
| 19.68 | — | — | — | — | 7.5448 | 7.212 |
| 17.22 | — | — | 11.003 | 6.987 | 5.9990 | 7.072 |
| 14.76 | — | — | 6.7289 | 7.358 | 6.2920 | 6.810 |
| 12.30 | — | — | 5.8009 | 6.706 | 4.2397 | 6.237 |
| 9.84 | 7.3692 | 4.759 | 5.2100 | 6.536 | 3.2048 | 5.400 |
| 7.38 | 5.5898 | 4.440 | 3.6845 | 4.861 | 3.4596 | 4.245 |
| 4.92 | 3.6745 | 3.703 | 2.2597 | 3.797 | 1.5553 | 2.953 |
| 2.46 | 1.5137 | 1.409 | 0.8520 | 1.256 | 1.1211 | 1.0347 |
| Note: — indicates $a$ is too large for given SWNT configuration, hence $G_c$ is not calculated | | | | | | |

Finally, the fracture energies of other brittle materials are given in Table 5 for comparison. Zigzag carbon nanotubes may be stated to be more brittle than graphite but tougher than Silicon.

**Table 5 Fracture energy of brittle materials [54-56]**

| Material | Fracture resistance, $G_c$ (J/m$^2$) |
|---|---|
| $Al_2O_3$ | 25~54 |
| Graphite | 85 (E=12 GPa), 68.1 (E=14.7 GPa) |
| SiC | 15.2~25.6 |
| Si | 2.7 |
| Si | 1.2 |
| NaCl | 0.25 |
| MgO | 1.2 |
| Zigzag tube (300 K) | 0.5 ~ 6.0 |

## 5. Conclusion

The fracture resistance of zigzag carbon nanotubes are investigated quantitatively in this study, by applying fracture mechanic concepts to nanostructure and modeling the deformation with atomistic simulation. The SWNTs' fracture resistance in the form of critical strain energy release rate $G_c$ is determined for (17,0), (28,0) and (35,0) SWNTs (each with aspect ratio 4) with cracks up to 29.5Å long. The fracture resistance is comparable with the fracture toughness of graphite and Silicon. A significant dependence of $G_c$ on crack length is observed: $G_c$ increases with crack length at small length, and tends to reach a constant value. The $G_c$ vs. crack length relation seems not to depend much on the diameter of the nanotubes. This behavior is strongly evocative of the rising R curve behavior for metals; we suspect that lattice trapping effect plays the role of crack-tip plasticity at the atomic level.

Dependence of $G_c$ on temperature is also investigated for SWNTs with aspect ratio 4. $G_c$ drops sharply as temperature increases; but for a given temperature the shape of the $G_c$ vs. crack length curve remains qualitatively the same.

The length effect on fracture resistance is investigated in a limited way: $G_c$ is computed only at 1K for the (17,0), (28,0) and (35,0) SWNTs each with aspect ratio 8. The fracture resistance shows no definite trend although in a majority of the cases the longer nanotubes are found to have higher fracture resistance than the shorter tubes. This difference seems to decrease as the tube diameter increases.

Future investigation should include fracture resistance of armchair and chiral SWNTs and MWNTs and effect of strain rates.

## 6. Acknowledgment

USA.

## 7. Reference

[1]     Iijima, S., *Helical microtubules of graphitic carbon.* Nature, 1991. **354**: p. 56.

[2]     Iijima, S., T. Ichihashi, and Y. Ando, *Pentagons, heptagons and negative curvature in graphite microtubule growth.* Nature, 1992. **356**(776).

[3]     Dresselhaus, M.S. and H. Dai, *Carbon nanotubes: Continued innovations and challenges.* Mrs Bulletin, 2004. **29**(4): p. 237-239.

[4]     Terrones, M., *Science and technology of the twenty-first century: Synthesis, properties, and applications of carbon nantoubes.* Annual Review of Materials Research, 2003. **33**: p. 419-501.

[5]     Treacy, M.M., et al., *Exceptional high Young's modulus observed for individual carbon nanotubes.* Nature, 1996. **381**(6584): p. 678.

[6]     Krishnan, A., et al., *Young's modulus of single-walled nanotubes.* Physical Review B, 1998. **58**(20): p. 14013-14019.

[7]     Poncharal, P., et al., *Electrostatic deflections and electromechanical resonances of carbon nanotubes.* Science, 1999. **283**(5047): p. 1513-1516.

[8]     Cuenot, S., et al., *Measurement of elastic modulus of nanotubes by resonant contact atomic force microscopy.* Journal of Applied Physics, 2003. **93**(9): p. 5650-5655.

[9]     Yu, M.F., O. Lourie, and M.J. Dyer, *Strength and breaking mechanism of multiwalled carbon nanotubes under tensile load.* Science, 2000. **287**(5453): p. 637-640.

[10]    Yu, M.F., et al., *Tensile loading of ropes of single wall carbon nanotubes and their mechanical properties.* Physical Review Letters, 2000. **84**(24): p. 5552-5555.

[11]    Wong, E.W., P.E. Sheehan, and C.M. Lieber, *Nanobeam mechanics: Elasticity, strength, and toughness of nanorods and nanotubes.* Science, 1997. **277**(5334): p. 1971-1975.

[12]    Salvetat, J.P., et al., *Elastic modulus of ordered and disordered multiwalled carbon nanotubes.* Advanced Materials, 1999. **11**(2): p. 161-165.

[13]    Salvetat, J.P., et al., *Elastic and shear moduli of single-walled carbon nanotube ropes.* Physical Review Letters, 1999. **82**(5): p. 944-947.

[14]    Demczyk, B.G., et al., *Direct mechanical measurement of the tensile strength and elastic modulus of multiwalled carbon nanotubes.* Materials Science and Engineering A, 2002. **334**(1-2): p. 173-178.

[15]    Rafii-Tabar, H., *Computational modelling of thermo-mechanial and transport peroperties of carbon nanotubes.* Physics Reports, 2004. **390**: p. 235-452.

[16]    Yu, M.F., T. Kowalewski, and R.S. Ruoff, *Investigation of the radial deformability of individual carbon nanotubes under controlled indentation force.* Physical Review Letters, 2000. **85**(7): p. 1456-1459.

[17]    Shen, W.D., et al., *Investigation of the radial compression of carbon nanotubes with a scanning probe microscope.* Physical Review Letters, 2000. **84**(16): p. 3634-3637.

[18]    Chesnokov, S.A., et al., *Mechanical energy storage in carbon nanotube springs.* Physical Review Letters, 1999. **82**(2): p. 343-346.

[19]    Tang, J., et al., *Compressibility and polygonization of single-walled carbon nanotubes under hydrostatic pressure.* Physical Review Letters, 2000. **85**(9): p. 1887-1889.

[20]    Sharma, S.M., et al., *Pressure-induced phase transformation and structural resilience of single-wall carbon nanotube bundles.* Physical Review B, 2001. **6320**(20): p. 205417.

[21]    Yu, M.F., et al., *Locked twist in multiwalled carbon-nanotube ribbons.* Physical Review B, 2001.

**64**(24): p. 241403.

[22]    Chopra, N.G., et al., *Fully collapsed carbon nanotubes.* 1995. **377**(6545): p. 135-138.

[23]    Falvo, M.R., et al., *Bending and buckling of carbon nanotubes under large strain.* Nature, 1997. **389**: p. 582.

[24]    Iijima, S., et al., *Structural flexibility of carbon nanotubes.* The Journal of Chemical Physics, 1996. **104**(5): p. 2089-2092.

[25]    Falvo, M.R., et al., *Bending and buckling of carbon nanotubes under large strain.* Nature, 1997. **389**(6651): p. 582-584.

[26]    Yu, M.F., B.I. Yakobson, and R.S. Ruoff, *Controlled sliding and pullout of nested shells in individual multiwalled carbon nanotubes.* Journal of Physical Chemistry B, 2000. **104**(37): p. 8764-8767.

[27]    Cumings, J. and A. Zettl, *Low-Friction Nanoscale Linear Bearing Realized from Multiwall Carbon Nanotubes.* Science, 2000. **289**(5479): p. 602-604.

[28]    Fennimore, A.M., et al., *Rotational actuators based on carbon nanotubes.* Nature, 2003. **424**(6947): p. 408-410.

[29]    Troiani, H.E., et al., *Direct observation of the mechanical properties of single-walled carbon nanotubes and their junctions at the atomic level.* Nano Letters, 2003. **3**(6): p. 751-755.

[30]    Marques, M.A.L., et al., *On the breaking of carbon nanotubes under tension.* Nano Letters, 2004. **4**(5): p. 811-815.

[31]    Nardelli, M.B., B.I. Yakobson, and J. Bernholc, *Mechanism of strain release in carbon nanotube.* Physical review B, 1998. **57**(8): p. R4277.

[32]    Dumitrica, T., T. Belytschko, and B.I. Yakobson, *Bond-breaking bifurcation states in carbon nanotube fracture.* Journal of Chemical Physics, 2003. **118**(21): p. 9485-9488.

[33]    Troya, D., S.L. Mielke, and G.C. Schatz, *Carbon nanotube fracture - differences between quantum mechanical mechanisms and those of empirical potentials.* Chemical Physics Letters, 2003. **382**(1-2): p. 133-141.

[34]    Yakobson, B.I., *Mechanical relaxation and intramolecular plasticity in carbon nanotubes.* Applied Physics Letters, 1998. **72**(8): p. 918.

[35]    Liew, K.M., X.Q. He, and C.H. Wong, *On the study of elastic and plastic properties of multi-walled carbon nanotubes under axial tension using molecular dynamics simulation.* Acta Materialia, 2004. **52**(9): p. 2521-2527.

[36]    Jiang, H., et al., *Defect nucleation in carbon nanotubes under tension and torsion: Stone-Wales transformation.* Computer Methods in Applied Mechanics and Engineering, 2004. **193**(30-32): p. 3419-3429.

[37]    Banhart, F., *Irradiation effects in carbon nanostructures.* Reports on Progress in Physics, 1999. **62**(8): p. 1181-1221.

[38]    Zhou, O., et al., *Defects in Carbon Nanostructures.* Science, 1994. **263**(5154): p. 1744-1747.

[39]    Charlier, J.C., *Defects in Carbon nanotube.* Accounts of Chemical Research, 2002. **35**: p. 1063-1069.

[40]    Saether, E., *Transverse mechanical properties of carbon nanotube crystals. Part II: sensitivity to lattice distortions.* Composite Science and Technology, 2003. **63**: p. 1551-1559.

[41]    Ouyang, M., et al., *Atomically resolved single-walled carbon nanotube intramolecular junctions.* Science, 2001. **291**: p. 97.

[42]    Lu, Q. and B. Bhattacharya, *Effect of randomly occurring Stone-Wales defects on mechanical properties of carbon nanotubes using atomistic simulation.* Nanotechnology, 2005(4): p. 555-566.

[43]     Lu, Q. and B. Bhattacharya. *Analysis of Randomness in Mechanical Properties of Carbon Nanotubes Through Atomistic Simulation". in 46th AIAA/ASME/ASCE/AHS/ASC Structures, Structural Dynamics & Materials Conference.* 2005. Austin, Texas.

[44]     Belytschko, T., et al., *Atomistic simulations of nanotube fracture.* Physical Review B, 2002. **65**: p. 235430.

[45]     Brenner, D.W., *Empirical potential for hydrocarbons for use in simulating the chemical vapor deposition of diamond films.* Physical Review B, 1990. **42**(15): p. 9458-9471.

[46]     Huhtala, M., et al., *Improved mechanical load transfer between shells of multiwalled carbon nanotubes.* Physical Review B, 2004. **70**(4).

[47]     Al-Jishi, R. and G. Dresselhaus, *Lattice-Dynamical Model for Graphite.* Physical Review B, 1982. **26**(8): p. 4514-4522.

[48]     Lu, Q. and B. Bhattacharya, *The role of atomistic simulations in probing the small-scale aspects of fracture - a case study on a single-walled carbon nanotube.* Engineering Fracture Mechanics, 2005(In Press).

[49]     Li, Q.M., *Dissipative flow model based on dissipative surface and irreversible thermodynamics.* Archive of Applied Mechanics, Springer-Verlag, 1999. **69**: p. 379-392.

[50]     Inglis, C.E., *Stresses in a plate due to the presence of cracks and sharp corners.* Transactions of the Institute of Naval Architects, 1913. **55**: p. 219-241.

[51]     Griffith, A.A., *The phenomena of rupture and flow in solids.* Philosophical Transactions of the Royal Society of London, 1920. **221**: p. 163-197.

[52]     Irwin, G.R., *Onset of Fast Crack Propagation in High Strength Steel and Aluminum Alloys.* Sagamore Research  Conference Proceedings, 1956. **2**: p. 289-305.

[53]     Anderson, T.L., *Fracture Mechanics: Fundamentals and Applications.* 2005, Boca Raton: Taylor and Francis.

[54]     Dowling, N.E., *Mechanical behavior of materials : Engineering methods for deformation, fracture, and fatigue.* 2nd ed. 1999, Upper Saddle River, NJ: Prentice Hall.

[55]     Munro, R.G., *Fracture toughness data for brittle materials.* NISTIR. 1998, Gaithersburg, Maryland: U.S. Dept. of Commerce, National Institute of Standards and Technology.

[56]     Wachtman, J.B., *Mechanical properties of ceramics.* 1996, New York: Wiley.

[57]     Lu, Q. and B. Bhattacharya. *Fracture resistance of single-walled carbon nanotubes through atomistic simulation. in 9th International Conference on Structural Safety and Reliability (ICOSSAR).* 2005. Rome, Italy: Millpress.